# Astro2020 Science White Paper

# Measuring Protostar Masses: The Key to Protostellar Evolution

**Thematic Areas:** ☐ Planetary Systems  ☒ Star and Planet Formation
☐ Formation and Evolution of Compact Objects  ☐ Cosmology and Fundamental Physics
☐ Stars and Stellar Evolution  ☐ Resolved Stellar Populations and their Environments
☐ Galaxy Evolution  ☐ Multi-Messenger Astronomy and Astrophysics

**Principal Author:**
Name: John J. Tobin
Institution: National Radio Astronomy Observatory
Email: jtobin@nrao.edu
Phone: 434-244-6815

**Co-authors:** (names and institutions)
Stella Offner (University of Texas), Patrick Sheehan (National Radio Astronomy Observatory), Zhi-Yun Li (University of Virginia), S. Tom Megeath (University of Toledo), Leslie Looney (University of Illinois), Nicole Karnath (University of Toledo), Joel Green (STSci), Rob Gutermuth (University of Massachusetts), Will Fischer (STSci), Ian Stephens (CfA/SAO), Michael M. Dunham (SUNY - Fredonia), Yao-Lun Yang (University of Texas)

**Abstract**:
   Knowledge of protostellar evolution has been revolutionized with the advent of surveys at near-infrared to submillimeter wavelengths. This has enabled the bolometric luminosities and bolometric temperatures (traditional protostellar evolution diagnostics) to be measured for large numbers of protostars. However, further progress is difficult without knowing the masses of the central protostars. Protostar masses can be most accurately determined via molecular line kinematics from millimeter interferometers (i.e., ALMA). Theoretical investigations have predicted the protostellar mass function (PMF) for various protostellar mass accretion models, and it is now imperative to observationally constrain its functional form. While ALMA has enabled protostellar mass measurements, samples approaching 100 sources are necessary to constrain the functional form of the PMF, and upgrades to ALMA and/or a new mm/cm facility will increase the feasibility of measuring such a large number of protostar masses. The masses of protostars will enable their stellar structure (radius and intrinsic luminosity), evolution, and accretion histories to be better understood. This is made more robust when effective temperatures and accretion rates can be measured via ground/space-based near to mid-infrared spectroscopy. Furthermore, access to supercomputing facilities is essential to fit the protostar masses via radiative transfer modeling and updated theoretical/numerical modeling of stellar structure may also be required.

Disks are a ubiquitous feature of the star and planet formation process and rotationally-supported disks are expected to form early in the star formation process via conservation of angular momentum during the gravitational collapse of the dense cloud core (Tobin et al. 2012; Terebey et al. 1984). Protostellar disks are important because most material that will eventually become incorporated into the star must pass through the disk in the protostellar phase and they are the progenitors of proto-planetary disks (e.g., Andrews et al. 2010). However, disks in the protostellar phase of star formation have not yet been afforded a broad characterization due to protostellar disks being embedded within dense envelopes of infalling gas and dust.

The protostellar disk is especially important because it enables the protostellar mass to be measured (e.g., Tobin et al. 2012; Murillo et al. 2013). The protostellar mass, the mass of the central stellar object, is a fundamental property that can only be determined for the youngest, deeply embedded protostars (known as Class 0 sources) through observations of their Keplerian disks or their resolved infalling envelopes on < 1000 au scales. Molecular line kinematics are needed to measure protostar masses because Class 0 protostars are typically embedded within infalling envelopes, extincting the protostar by 10s to 100s of $A_v$. More-evolved, Class I protostars, typically have less opaque envelopes and can, in some cases, have their spectral types determined through near-infrared spectroscopy (e.g., Connelley & Greene 2009). However, accurate evolutionary tracks are needed to infer a stellar mass from the spectral type.

Once the protostar mass is known, the implications of many other observable properties (i.e., luminosity, disk mass, envelope mass, bolometric temperature) can be put into their proper context. The distribution of protostar masses can shed light on the underlying physics of stellar mass assembly (McKee & Offner 2010), enable a better connection between the 'evolutionary stage' of a protostar to its 'observational class' (Dunham et al. 2014), and constrain protostellar evolutionary models (Hosokawa et al. 2011; Baraffe et al. 2011). Furthermore, kinematic mass measurements of more-evolved young stars without envelopes offer the opportunity to obtain direct spectral and luminosity information and a direct measurement of stellar mass from the proto-planetary disk rotation. This enables crucial tests of stellar evolution models and reduces the ambiguity in the age of a pre-main sequence star (Sheehan et al. 2019), which have previously required measurements of eclipsing binaries (Stassun et al. 2014).

**Measuring Protostar Masses**

The accurate measurement of protostar masses requires both sensitive spectrally-resolved molecular line observations and high-spatial resolution. Protostar mass measurements are best provided by observations of molecular lines tracing emission from the rotationally-supported protostellar disk, requiring high enough spatial resolution to resolve the disk having a typical radius of ~25 au in the protostellar phase (Tobin et al. 2019 in prep). But, even when the disk is not well-resolved, the line emission from the rotating/infalling inner envelope, combined with unresolved (or marginally-resolved) emission from the disk, enables the central protostar mass to be constrained. The Atacama Large Millimeter/submillimeter Array (ALMA) enables the use protostellar disk and inner envelope kinematics to measure the masses of protostars in the Class 0 and Class I phases, in addition to proto-planetary disks. Prior to ALMA, measuring protostar/pre-main sequence star masses was possible for some protostellar sources (Tobin et al. 2012; Harsono et al. 2014), but it was challenging to obtain enough sensitivity to map the disk rotation curves and inner envelope material with high enough S/N.

The most reliable tracers of inner envelope and disk kinematics for low-mass protostars are

$^{13}$CO, C$^{18}$O, and C$^{17}$O, because they trace progressively higher column densities along the line of sight and CO is pervasive in the interstellar medium. Both $^{13}$CO and C$^{18}$O, can be good disk/envelope tracing molecules for Class I protostars, while C$^{18}$O and C$^{17}$O are more favorable for Class 0 protostars because of the higher column densities of the surrounding envelopes. Disks in the protostellar phase also tend to be warmer than proto-planetary disks due to backwarming from the envelope (van 't Hoff et al. 2018), so CO molecules tend to be in the gas phase on scales < 1000 au. Also, luminosity bursts seem to play a role in keeping CO in the gas phase within 1000 au (Jorgensen et al. 2015; Frimann et al. 2017). Other molecules can also trace disks/inner envelopes, (e.g., SO, H$_2$CO; Sakai et al. 2014), but not with the reliability of CO.

Protostellar mass measurements are now feasible for small samples of protostars with ALMA. However, the observing time needed to detect these molecules with high S/N is expensive, even with ALMA. Thus, there are still only about ~20 protostars with robust mass measurements from ALMA and they do not constitute an unbiased sample since the median mass of the protostars is ~0.6 M$_{sun}$ (Yen et al. 2017). Furthermore, protostellar disks can have large enough mass such that the dust continuum emission is optically thick at 1.3 mm, where the most convenient $^{13}$CO and C$^{18}$O (2-1) transitions are found. Thus, it may be favorable to observe the CO (1-0) transitions at ~3 mm, but the $^{13}$CO and C$^{18}$O lines will be intrinsically fainter in flux density (>= 4x fainter depending on line opacity) at 3 mm vs 1.3 mm for equivalent brightness temperature. Therefore, it will be difficult to use the $^{13}$CO and C$^{18}$O (1-0) lines to map the molecular line kinematics with ALMA. A new or upgraded facility with the ability to observe at 3mm is required to observe the CO (1-0) isotopologue lines from disks at longer wavelengths were dust is more transparent.

With sufficiently high-S/N molecular line data, there are several methods that can be used to measure the protostellar masses. Early efforts used Gaussian fits to each channel of a spectral cube under the assumption that the high-velocity, point-like emission directly mapped to a radius (Tobin et al. 2012; Yen et al. 2013; Ohashi et al. 2014). This assumption breaks down once the disk is sufficiently resolved and these simple methods can introduce a systematic error toward higher masses. Another method, proven with numerical simulations and radiative transfer, fits the limits of a position-velocity diagram (Seifried et al. 2016) and has been utilized with observational data (Ginsburg et al. 2018). The latter method is probably the most reliable of the observation-only determinations. Newer methods have been developed using Markov Chain Monte Carlo (MCMC) parameter sampling and molecular line radiative transfer models to fit models in the uv-plane, utilizing the full spatial dynamic range of the data (Czekala et al. 2016; Sheehan et al. 2019). These models can include both a disk and a rotating/infalling envelope to enable mass measurements with molecular line observations that do not fully resolve the disk kinematics. MCMC methods are proving their utility, but can be computationally expensive. Thus, access to high-performance computational facilities are essential for measuring protostar masses.

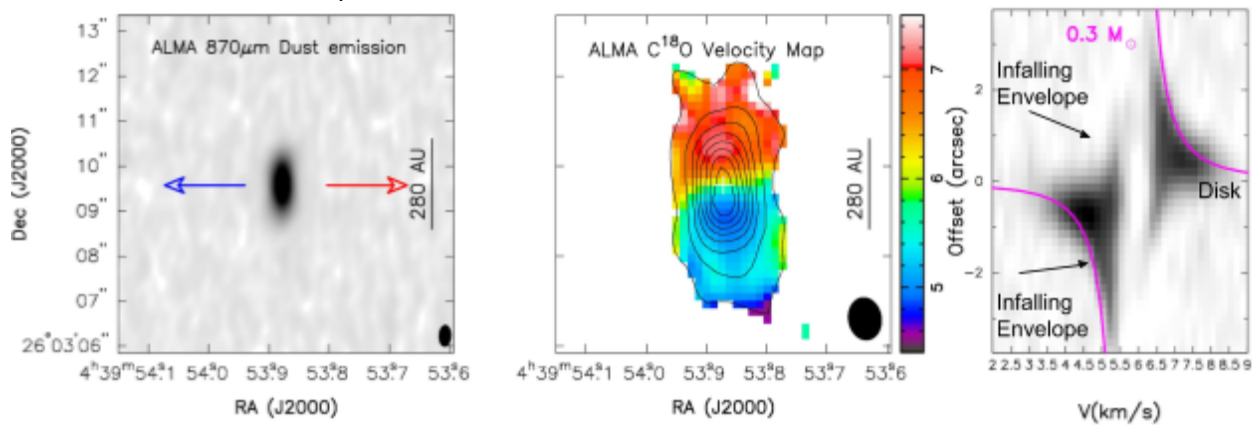

Figure 1 - L1527 IRS, the first extremely young protostar to have a Keplerian disk identified (Tobin et al. 2012, Ohashi et al. 201; Aso et al. 2017), observed in ALMA dust continuum (left), C$^{18}$O (J=2-1) molecular line (middle), and C$^{18}$O position-velocity diagram (2D spectrum) (right). The dust continuum shows the resolved structure of the disk (size and mass); the C$^{18}$O line center velocity map (color scale) shows the rotation signature of the disk. The position-velocity diagram (right) along the major axis of the disk shows the rotation velocities, indicating a 0.3 M$_{sun}$ protostar (magenta line), in addition to the infalling envelope beyond the disk (> ~1"), which also can enable protostar mass constraints when resolution is not high enough to fully-resolve the disk. See also Tobin et al. (2019) for intermediate resolution data.

**The Protostellar Mass Function**

The measurement of protostellar masses for statistical samples of protostars will enable, for the first time, the protostellar mass function (PMF) to be directly measured. The PMF is the instantaneous mass distribution of the central protostars, and the PMF should be time independent in the limit of constant star formation rate (over ~1 Myr). Using various theoretical prescriptions for protostellar mass accretion, McKee & Offner (2010) predicted possible functional forms of the PMF (Figure 2). Attempts have been made to connect this to observations using the predicted luminosity functions (Offner & McKee 2011; Dunham et al. 2014), but the underlying mass function cannot be uniquely constrained due to degeneracy between mass and accretion rate. Furthermore, the fraction of accretion luminosity is unknown through the protostellar phase and strongly depends on age and protostar mass. Thus, the PMF can only be directly constrained by measuring protostar masses.

The theoretical PMFs shown in Figure 2 consider a single star-forming region. Thus, it is most ideal to observe the PMF for a single star forming region. Multiple star forming regions could be combined, but this may result in integrating over different initial conditions. The only nearby star-forming region with enough protostars to measure the mass function is Orion, hosting at least 279 protostars that have detectable disk emission in dust continuum with ALMA at 0.1" resolution (Tobin et al. 2019 in prep.); ~100 of this sample are Class 0 protostars. Thus, ALMA can enable the PMF to be observed for the first time in a single star-forming region if a sufficient amount of observing time is invested (1-2 hours per source). However, a significant upgrade to the spectral line sensitivity of ALMA and/or a new facility operating at 3mm could greatly increase the number of protostars with measured masses to populate the PMF and enable investigations into its possible regional variation. For such a large sample, Science Ready Data Products from the observatories will be essential. Also, *Gaia* and the Gould Belt distance survey with the Very Long Baseline Array have provided accurate distances to star forming regions within 500 pc, which are essential for measuring masses from disk kinematics because they depend on the physical size of the observed region (e.g., Kounkel et al. 2018).

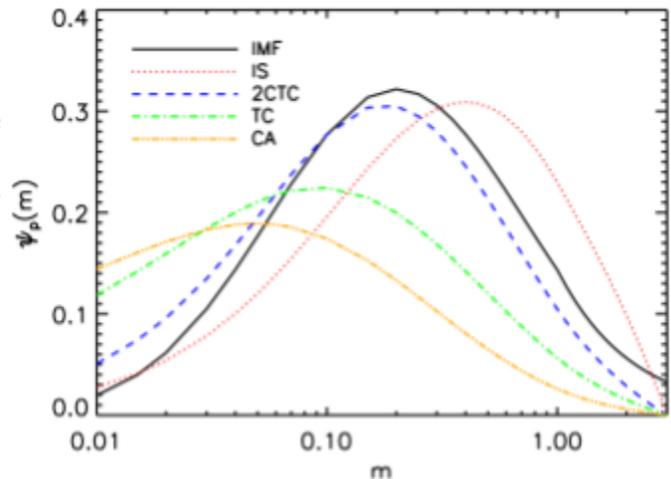

**Figure 2.** Predicted protostellar mass functions for a single star forming region with protostellar collapse preceeding by different physical processes. IMF- Chabrier initial mass function, IS - Isothermal Sphere, 2CTC - Two component Turbulent Core, TC - Turbulent Core, CA - Competitive Accretion, reproduced from Mckee & Offner (2010).

In order to characterize the PMF sufficiently well, uncertainties in each mass bin at the peak of the PMF should be less than 20%. Assuming that a mid to far-infrared selected sample of protostars spans the entire PMF and that the PMF is described by the Chabrier (2005) IMF, about 30% of protostars will have M < 0.1 $M_{sun}$, and 70% with M > 0.1 $M_{sun}$. If we assume that protostars with masses > 0.1 $M_{sun}$ can be feasibly measured to 20% accuracy, then a sample of ~140 protostars is required to characterize both the peak and the functional form of the PMF with 5 logarithmically spaced bins between 0.1 $M_{sun}$ and 1.77 $M_{sun}$. This will enable the analytical prescriptions for mass accretion to be constrained or ruled-out. Furthermore, a measured PMF will provide a valuable reference point for simulations of the multi-scale star formation process (e.g., Bate 2018; Krumholz et al. 2016).

**Connecting Protostellar Evolution to Protostar Masses**

Protostar masses are a key uncertainty in the understanding early protostellar evolution. The standard diagnostic that has been developed and extended with increasing samples over the past two decades is the bolometric luminosity ($L_{bol}$) vs. bolometric temperature ($T_{bol}$) diagram or BLT diagram (Myers & Ladd 1993; Fischer et al. 2017). $L_{bol}$ and $T_{bol}$ are observable properties can be determined with broad wavelength coverage of spectral energy distributions (SEDs) from near-infrared to millimeter wavelengths, and it is used to classify young stellar objects as Class 0, Class I, and Class II and examine their evolution. With the progress enabled by *Spitzer* and *Herschel* surveys, the BLT diagram is well-populated for star-forming regions within 500 pc (see WP by Green et al. and Fischer et al.). However, despite extensive theoretical study (Young & Evans 2005; Dunham et al. 2010; Fischer et al. 2017) the BLT diagram has substantial scatter owing to either the protostar masses or variable accretion luminosity, but could indicate an overall exponentially declining mass accretion rate from Class 0 to Class I.

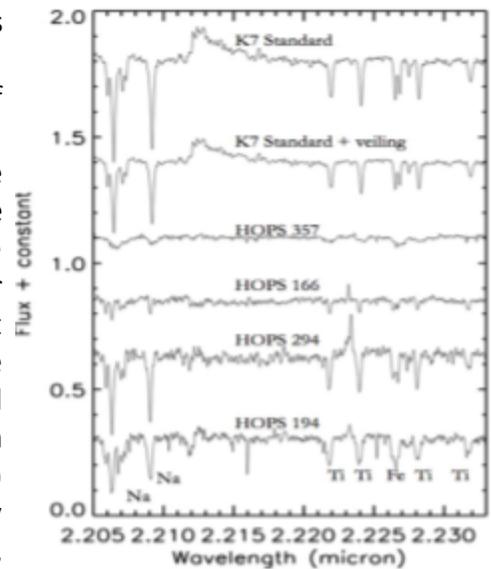

Measuring the masses of a significant number of protostars would enable a connection between luminosity, evolutionary state, and protostar mass. This will enable the ratio of accretion luminosity to protostellar luminosity to be determined for a large sample in order to understand the rate of stellar mass assembly throughout protostellar evolution. Protostar masses will also enable significant expansion of our knowledge of early stellar structure. The more-evolved protostars that can have near-infrared spectra observed (and some Class 0 protostars with enough invested time Greene et al. 2018) enable the relationship between mass and spectral type to be characterized. Highly efficient spectrographs, using immersion gratings (i.e., IGRINS), are making near-infrared spectroscopy at high-resolution feasible for spectral typing protostars in the presence of significant absorption line veiling.

The protostar luminosity can be disentangled from accretion luminosity with measurements of near-infrared hydrogen recombination lines (ground-based lines and others in the mid-infrared by JWST) that are correlated with accretion luminosity (e.g., Muzerolle et al. 1998). The protostellar radius can then be estimated using the knowledge of effective temperature and stellar luminosity. The combined utilization of millimeter-wave spectroscopy, infrared spectroscopy, near-infrared to millimeter photometry, will enable the mass vs. radius relationship of protostars to be known for the first time. This can be accomplished with a combination of archival data (2MASS, *Spitzer, Herschel*), current ground-based telescopes >3m with near-infrared spectrographs (high, medium, and low dispersion), JWST, and ALMA. However, this work will be enhanced with an ELT to enable observations of more deeply embedded protostars.

**Figure 3.** Near-infrared spectra toward Class I protostars (and standards) taken with IGRINS from a single order. The high spectral resolution enables narrow absorption lines to be revealed, highlighting the utility of high-resolution infrared spectrographs to measure spectral types of protostars.

With these constraints on protostar masses, protostar luminosities, accretion luminosities, and perhaps stellar radii, it will also be possible to unravel the evolution of accretion in protostars. The accretion luminosity for protostars without spectral types can be estimated using the protostar mass and stellar structure models to infer the expected luminosity of the protostar. When connected to the BLT properties of protostars, it can be determined if accretion is a smoothly declining process or if it is a stochastic process characterized by large variations in the protostellar phase. Note that time domain observations of protostars will approach this problem from a complementary direction (WPs by W. Fischer, J. Green, and T. Hunter) and will most importantly constrain the timescale for accretion variability and the level of variation over short time periods. However it is presently unclear if current stellar evolution models are accurate for protostars and additional theoretical work on early stellar structure and the birthline may be required.

**Pre-Main Sequence Stellar Masses and Evolution**

Measurement of protostellar masses and masses of pre-main sequence (pre-MS) stars can calibrate evolutionary tracks, which, despite significant improvements over the past decades, show significant with deviations from observations (e.g. Stassun et al. 2014, Bell 2016). This is in part due to the limited sample of pre-MS stars with well-measured masses; currently only ~20 young eclipsing binaries have reliable mass measurements (e.g. Stassun et al. 2014). Masses and radii determined for late stage protostars will constrain the initial radii of pre-MS contraction, a primary uncertainty in models, and test models predicting that the initial radii depend on the accretion history (Baraffe et al. 2012). Young stars are commonly surrounded by disks (e.g. Hernandez et al. 2008); if stellar masses can be accurately measured with high precision from disk kinematics, then these sources would provide strong constraints on evolutionary tracks. This route is particularly promising because Gaia is now measuring precise distances to nearby pre-MS stars, thereby removing a significant source of uncertainty in stellar mass measurements (e.g., Czekala et al. 2015, Sheehan et al. 2019). The calibrated tracks will be used to convert the location of pre-MS stars in HR diagrams into masses and ages, enabling more reliable determinations of IMFs, star formation histories, and the timescales for disk evolution.

**Summary**

The potential to reveal the protostellar mass function and better understand the evolution of protostars in the next decade will be a major advance enabled by ALMA, utilized in conjunction with near-to-mid-infrared facilities and possibly a new mm/cm-wave facility. Protostar mass is the key to understanding protostellar evolution in the context of multi-wavelength observations. The combined constraints provided by near-infrared to submillimeter photometry and spectroscopy at infrared and sub/millimeter wavelengths in the next decade will enable the structure and evolution of protostars to be understood in ways not currently possible in the absence of large samples of protostar masses.

**References**
Baraffe, I., Vorobyov, E., & Chabrier, G. et al. 2012, ApJ, 756, 118
Bate, M. 2018, MNRAS, 475, 5618
Bell, C. 2016, Cool Stars 19, arXiv:1609.03209,
Chabrier, G. 2005, ASSL, 327, 41
Connelley, M. & Greene, T. 2009, AJ, 140 1214
Czekala, I. et al. 2015, ApJ, 806, 154
Dunham, M. et al. 2010, ApJ, 710, 470
Dunham, M. et al. 2014, PPVI, 195
Fischer, W. J. et al. 2017, ApJ, 840, 69
Frimann, S. et al. 2017, A&A, 602, 120
Ginsburg, A. et al. 2018, ApJ, 860, 19
Greene, T. et al. 2018, ApJ, 862, 85
Hennebelle, P. et al. 2016, ApJL, 830, 8
Hernandez, J. et al. 2008, ApJ, 686, 1195
Hosokawa, T. et al. 2011, 738, 140
Li, Z.-Y. et al. 2014, PPVI, 173
Jorgensen, J. et al. 2015, A&A, 579, 23
Krumholz, M. et al. 2016, ApJ, MNRAS, 460, 3272
McKee, C. & Offner, S. 2010, ApJ, 716, 167
Myers, P. C. & Ladd, E. 1993, ApJ, 413, 47
Murillo, N. et al. 2013, A&A, 560, 103
Muzerolle, J. et al. 1998, ApJ, 116, 2965
Offner, S. & McKee, C. 2011, ApJ, 736, 53
Ohashi, N. et al. 2014, ApJ, 796, 131
Sakai, N. et al. 2014, Nature, 507, 78
Seifried, D. et al. 2016, MNRAS, 459, 1892
Sheehan, P. et al. 2019, ApJ, in press., arXiv:1903.00032
Stassun, K. et al. 2014, ApJ, New Astronomy Reviews, 60, 1
Terebey, S. et al. 1984, ApJ, 286, 529
Tobin, J. J., et al. 2019, ApJ, in preparation.
Tobin, J. J. et al. 2012, Nature, 492, 83
van 't Hoff, M. et al. 2018, A&A, 615, 83
Yen, H.-W. et al. 2013, ApJ, 772, 22
Young, C. & Evans, N. 2005, ApJ, 627, 293